\documentclass[aps,prb,twocolumn,groupedaddress]{revtex4-1}

\usepackage{amsmath,amssymb}
\usepackage{bm}
\usepackage{graphicx}
\usepackage{ascmac}

\usepackage{amsmath, amssymb}
\usepackage{type1cm}

\begin{document}

\title{Impact of graphene quantum capacitance on transport spectroscopy}

\author{K. Takase}
\email{takase.keiko@lab.ntt.co.jp}
\affiliation{NTT Basic Research Laboratories, NTT Corporation, 3-1 Morinosato-Wakamiya, Atsugi 243-0198, Japan 
}
\author{S. Tanabe}
\affiliation{NTT Basic Research Laboratories, NTT Corporation, 3-1 Morinosato-Wakamiya, Atsugi 243-0198, Japan 
}
\author{S. Sasaki}
\affiliation{NTT Basic Research Laboratories, NTT Corporation, 3-1 Morinosato-Wakamiya, Atsugi 243-0198, Japan 
}
\author{H. Hibino}
\affiliation{NTT Basic Research Laboratories, NTT Corporation, 3-1 Morinosato-Wakamiya, Atsugi 243-0198, Japan 
}
\author{K. Muraki}
\affiliation{NTT Basic Research Laboratories, NTT Corporation, 3-1 Morinosato-Wakamiya, Atsugi 243-0198, Japan 
}

\date{\today}

\begin{abstract}

We demonstrate experimentally that graphene quantum capacitance $C_{\mathrm{q}}$ can have a strong impact on transport spectroscopy through the interplay with nearby charge reservoirs. The effect is elucidated in a field-effect-gated epitaxial
graphene device, in which interface states serve as charge reservoirs. The Fermi-level dependence of $C_{\mathrm{q}}$ is manifested as an unusual parabolic gate voltage ($V_{\mathrm{g}}$) dependence of the carrier density, centered on the
Dirac point. Consequently, in high magnetic fields $B$, the spectroscopy of longitudinal resistance ($R_{xx}$) vs.~$V_{\mathrm{g}}$ represents the structure of the unequally spaced relativistic graphene Landau levels (LLs).
$R_{xx}$ mapping vs.~$V_{\mathrm{g}}$ and $B$ thus reveals the vital role of the zero-energy LL on the development of the anomalously wide $\nu=2$ quantum Hall state.

\end{abstract}


\maketitle

\section{Introduction}

Graphene has been attracting great interest, much of which is linked to the
behavior of its charge carriers that mimic massless Dirac fermions as a result
of the relativistic band structure referred to as Dirac cones. \cite{Novoselov, Novoselov2,Zhang} Among various
experimental techniques designed to probe Dirac cones, the measurement of quantum
capacitance $C_{\mathrm{q}}$, \cite{Chen,XiaNatureNanotech,Ponomarenko,Ensslin,XiaAPL} defined as $C_{\mathrm{q}}=e^{2}(\mathrm{d}n_{\mathrm{G}%
}/\mathrm{d}\varepsilon_{\mathrm{F}})$ with $n_{\mathrm{G}}$ the charge
carrier density and $\varepsilon_{\mathrm{F}}$ the Fermi level ($e$%
:~elementary charge), is unique in that it directly probes the density of
states (DOS) at the Fermi level. The effects of $C_{\mathrm{q}}$ usually appear as
a minor modification to the measured capacitance, expressed as
$C_{\mathrm{meas}}=(1/C_{\mathrm{ox}}+1/C_{\mathrm{q}})^{-1}$ ($C_{\mathrm{ox}%
}$: capacitance of gate oxide), which thus becomes discernible only in the
vicinity of the charge neutrality point (Dirac point) where $1/C_{\mathrm{q}%
} \gtrsim 1/C_{\mathrm{ox}}$. \cite{Chen,XiaNatureNanotech,Ponomarenko,Ensslin,XiaAPL} Using a very thin ($10$~nm)
high dielectric gate insulator, Ponomarenko
\textit{et al}.~were able to measure graphene quantum capacitance with high
accuracy, which revealed the smearing of the Dirac cone by charge puddles and
the broadening of relativistic graphene Landau levels (LLs) in high magnetic
fields ($B$). \cite{Ponomarenko} These studies demonstrate that $C_{\mathrm{q}}$ indeed
depends on $\varepsilon_{\mathrm{F}}$ \cite{Chen,XiaNatureNanotech,Ponomarenko,Ensslin,XiaAPL} unlike in conventional
two-dimensional (2D) systems. \cite{Luryi} However, with the standard setup for
transport spectroscopy using a gate oxide a few $100$~nm thick, the
feature of graphene quantum capacitance reflecting the Dirac cone is
not manifested, because it is overridden by $C_{\mathrm{ox}}$; accordingly, the usual linear
relation between $n_{\mathrm{G}}$ and gate voltage $V_{\mathrm{g}}$ is routinely assumed as in
conventional 2D systems.

In this paper, we demonstrate that graphene quantum capacitance can have an impact on 
transport spectroscopy over a wide carrier density range away from the
Dirac point, through the interplay with nearby charge reservoirs. Here we
elucidate the effect in high-mobility field-effect-gated epitaxial graphene,
in which the role of charge reservoirs is played by interface states that
exist at the SiC substrate and in the gate insulator. 
The interplay between the $\varepsilon_\mathrm{F}$-dependent graphene quantum capacitance and the nearly energy-independent interface state DOS modifies the usual relation $n_\mathrm{G} \propto V_\mathrm{g}-V_\mathrm{D}$ ($V_\mathrm{D}$: gate voltage at the Dirac point) into a nontrivial one $\varepsilon_\mathrm{F} \propto V_\mathrm{g}-V_\mathrm{D}$ [i.e., $|n_\mathrm{G}| \propto$ ($V_\mathrm{g}-V_\mathrm{D}$)$^2$] when the interface states dominate.
In high $B$, transport spectroscopy vs.~$V_{\mathrm{g}}$ consequently emerges as a manifestation of the
energy spectrum of the unequally spaced relativistic graphene LLs. This
additionally allows us to deduce the energy width of the extended state in the
$N=0$ LL ($N$: the LL orbital index) from the width of the longitudinal resistance ($R_{xx}$) peak.
The mapping of $R_{xx}$ vs.~$V_{\mathrm{g}}$ and $B$ provides an intuitive picture
of the vital role of the zero-energy $N=0$ LL in the development of the
anomalously wide $\nu=2$ quantum Hall (QH) state, which has been commonly
observed in low-density epitaxial graphene. \cite{Jobst, Janssen, Lara-Avila, Wu, ShenJAP, Jouault}
We explain our data coherently by using a device model containing interface states. Our results
suggest that similar effects, although not as apparent, may be at work in
the transport spectroscopy of graphene including nano- and heterostructures.\cite{Tutuc}

\begin{figure}[htbp]
\includegraphics[ width=.8\linewidth ]{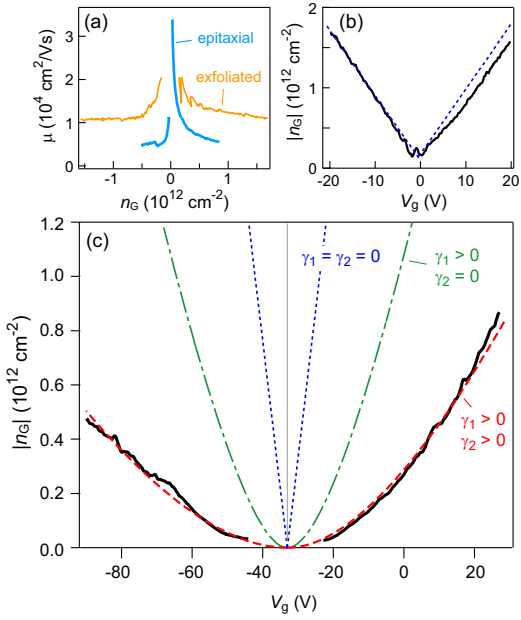} \label{fig1} 
\caption{ (Color online) (a) Mobility vs.~carrier density $n_\textrm{G}$ for epitaxial and exfoliated graphene devices. (b),(c) $|n_\textrm{G}|$ vs.~$V_{\mathrm{g}}$  for (b) exfoliated graphene and (c) epitaxial graphene. The dotted lines indicate the relation expected from each gate capacitance. The dash-dotted and dashed lines indicate calculations for different interface state densities: $\gamma_1 \rightarrow \infty$ and $\gamma_2 = 0$ (dash-dotted), $\gamma_1 = 5.0 \times 10^{12} \mathrm{eV}^{-1}\mathrm{cm}^{-2}$ and $\gamma_2 = 3.7$ $(4.8) \times 10^{13}$ $\mathrm{eV}^{-1}\mathrm{cm}^{-2}$ for $n_\textrm{G} > 0$ ($n_\textrm{G} < 0$) (dashed).}
\end{figure}

\section{Sample and method}

Our epitaxial graphene device was fabricated from graphene grown on 4H-SiC(0001) and has a top gate with a gate insulator made of HSQ/SiO$_2$ (120 nm/40 nm). \cite{TanabeAPEX} The device is a Hall-bar with width $W$ = 5 $\mu$m and length $L$ = 15 $\mu$m. 
For comparison, we also studied an exfoliated graphene device of a comparable size, fabricated on a Si/SiO$_2$ (285 nm) substrate using the conventional method. \cite{Novoselov} 
Transport measurements were performed at a temperature of 1.5 K using a standard lock-in technique with a current of less than 1 $\mu$A.
For epitaxial graphene, we have studied more than ten devices
fabricated from several different SiC/graphene wafers.
Here we present data taken from one sample for the consistency
of the analysis, but we emphasize that similar results are
obtained for all the samples studied.

\section{EXPERIMENT AND ANALYSIS}
\subsection{Results in zero and low magnetic fields}

Figure 1(a) compares the mobility $\mu$ [$= 1/n_\textrm{G} e R_\textit{xx} \times (L/W)$] of exfoliated and epitaxial graphene devices, plotted as a function of $n_\textrm{G}$ determined from the Hall coefficient in low fields ($\leq$ 0.5 T). 
The two devices have comparably high mobilities,
whereas a striking difference is observed for the $V_\mathrm{g}$ dependence of $|n_\textrm{G}|$.
Exfoliated graphene shows a normal $|n_\textrm{G}| \propto V_\mathrm{g}$ relation with a slope that is in accordance with the gate capacitance [Fig.~1(b)].
In contrast, the change in $|n_\textrm{G}|$ with $V_\mathrm{g}$ in epitaxial graphene is much smaller than that expected from the gate capacitance, $|n_\textrm{G}| = C_\mathrm{ox}|V_\mathrm{g} - V_\mathrm{D}|/e$ [dotted line in Fig.~1(c)]. More intriguingly, the dependence of $|n_\textrm{G}|$ on $V_\mathrm{g}$ is nonlinear, being parabolic-like around the Dirac point. 
The symmetry of the feature around the Dirac point implies that it arises from an intrinsic property of graphene.

The reduced action of $V_{\mathrm{g}}$ on $n_{\mathrm{G}}$ indicates the
existence of carrier trap states that accommodate a portion of the charges
induced by the gate. The observed parabolic dependence $\left\vert
n_{\mathrm{G}}\right\vert \varpropto(V_{\mathrm{g}}-V_{\mathrm{D}})^{2}$,
combined with $\left\vert n_{\mathrm{G}}\right\vert \varpropto\varepsilon
_{\mathrm{F}}^{2}$ specific to monolayer graphene, \cite{Novoselov2} suggests that in the present case
the unusual relation $\varepsilon_{\mathrm{F}}\varpropto V_{\mathrm{g}%
}-V_{\mathrm{D}}$ holds, as opposed to the normal behavior $n_{\mathrm{G}%
}\varpropto V_{\mathrm{g}}-V_{\mathrm{D}}$. As we show below, this can be
explained if the trap states have a large and nearly constant DOS.

\begin{figure}[tbp]
\includegraphics[ width=\linewidth ]{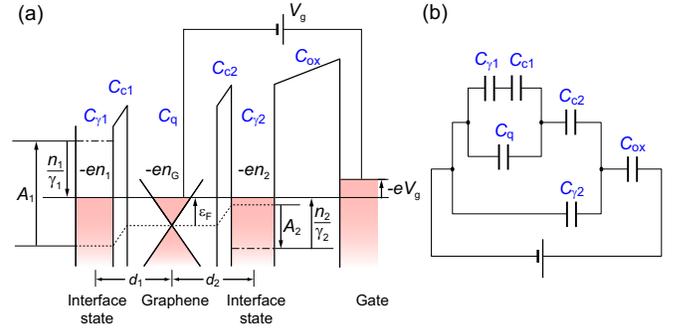} \label{fig2}
\caption{ (Color online) (a) Schematic energy diagram of a top-gated graphene device incorporating two kinds of interface states below and above graphene, with DOS $\gamma_1$ and $\gamma_2$. The charge density in the lower (upper) interface state is $-en_{1(2)}$. $A_{1(2)}$ is the work-function difference between graphene and the lower (upper) interface state. The figure represents a case where $A_1 > 0, A_2 < 0, n_1 < 0, n_2 > 0$, and $V_\mathrm{g} < 0$. (b) Equivalent circuit of (a).}
\end{figure}

\begin{figure*}[htb]
\includegraphics[ width=.95\linewidth ]{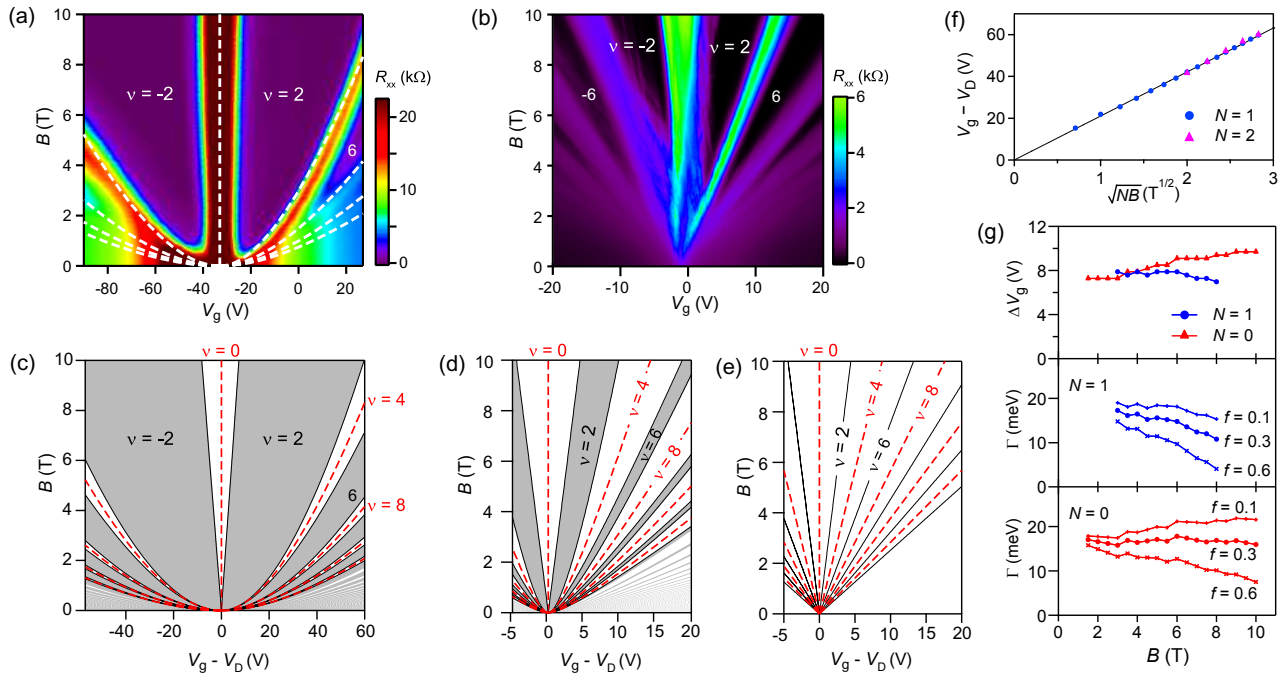} \label{fig3} 
\caption{(Color online) (a) $R_{xx}$ mapping vs.~$V_{\mathrm{g}}$ and $B$ for epitaxial
graphene. The white dashed lines, indicating the positions of half-filled LLs,
are calculated using the interface state densities $\gamma_{1}=5.0\times
10^{12}$~eV$^{-1}$cm$^{-2}$ and $\gamma_{2}=3.7$ ($4.8$) $\times10^{13}%
$~eV$^{-1}$cm$^{-2}$ for $n_{\mathrm{G}}>0$ ($n_{\mathrm{G}}<0$), obtained
from the low-field data in Fig.~1(c). (b) $R_{xx}$ mapping vs.~$V_{\mathrm{g}%
}$ and $B$ for exfoliated graphene. (c)-(e) Mapping of regions with integer
(gray) and non-integer (white) LL fillings versus $V_{\mathrm{g}}$ and $B$,
calculated using different interface state densities, (c) the same as in (a), (d)
$\gamma_{1}=5.0\times10^{12}$~eV$^{-1}$cm$^{-2}$ and $\gamma_{2}=4.8$
$\times10^{11}$~eV$^{-1}$cm$^{-2}$, and (e) \ $\gamma_{1}=\gamma_{2}=0$. The dashed lines represent half fillings of four-fold degenerate graphene LLs.
(f) $V_{\mathrm{g}}$ positions of $R_{xx}$ peaks in the data of (a) plotted
vs.~$\sqrt{NB}$ for the $N=1$ and $N=2$ LLs. (g) FWHM of $R_{xx}$ peaks and
the width $\Gamma$ of extended states deduced for different $f$ values, plotted
vs.~$B$ for the $N=0$ and $N=1$ LLs. $f$ is the fraction of the extended
states in the whole LL DOS.
}
\end{figure*}

\begin{figure}[htbp]
\includegraphics[ width=\linewidth ]{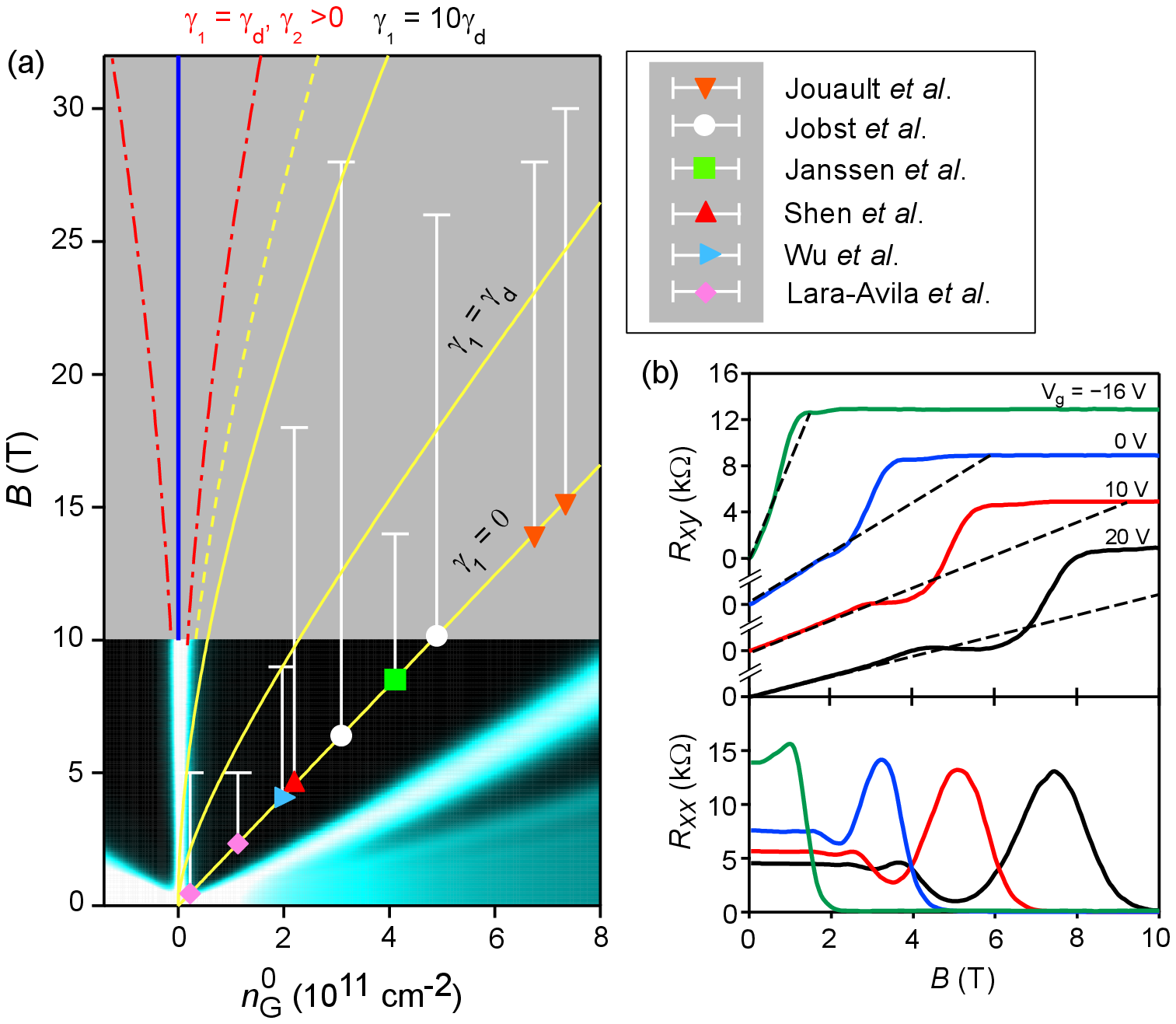} \label{fig4} 
\caption{(Color online) (a) Range of fields over which the $\nu=2$ QH regime was observed to
be extended to higher $B$ in previous reports, \cite{Jobst, Janssen, Lara-Avila, Wu, ShenJAP, Jouault} overlaid on our
$R_{xx}$ data replotted vs.~a low-field carrier density $n_{\mathrm{G}}^{0}$ and
$B$. 
Equation (3) calculated with $\gamma_{1}=0$,
$\gamma_{\mathrm{d}}$, and $10\gamma_{\mathrm{d}}$ ($\gamma_{\mathrm{d}}%
\equiv5.0\times10^{12}$~eV$^{-1}$cm$^{-2}$) with $\gamma_{2}=0$ are shown, with the
limit $\gamma_{1}\rightarrow\infty$ shown by the dashed line. Dash-dotted lines indicates
the calculation using $\gamma_{1}$ and $\gamma_{2}$ deduced for our sample.
(b) $R_{xy}$ and $R_{xx}$ versus magnetic fields for different $V_{\mathrm{g}%
}$. The dashed lines, linear extrapolations of $R_{xy}$ from low fields,
indicate the positions of $\nu=2$ expected from $n_{\mathrm{G}}^{0}$.} 
\end{figure}

We developed a model containing interface states below and above the graphene
layer, each of which has a constant DOS $\gamma_{1(2)}$ and
geometrical capacitance $C_{\mathrm{c}1(2)}=$ $\varepsilon_{1(2)}/d_{1(2)}$
with graphene, where $d_{1(2)}$ and $\varepsilon_{1(2)}$ are the distance and
effective permittivity between graphene and the interface states below (above)
[Fig.~2(a)]. The lower interface states represent the dangling-bond states at
the SiC substrate, \cite{Janssen,Kopilov,Varchon,Sonde} while the upper interface states represent charge traps in
the gate insulator. \cite{Sze} By noting that the $V_{\mathrm{g}}$-induced charge
redistribution between graphene and the interface states is influenced by both
the electrostatic potentials associated with $C_{\mathrm{c}1(2)}$ and the
chemical potentials associated with their quantum capacitances  given
by $C_{\mathrm{q}}=e^{2}(\mathrm{d}n_{\mathrm{G}}/\mathrm{d}\varepsilon
_{\mathrm{F}})$ \cite{Chen,XiaNatureNanotech,Ponomarenko,Ensslin,XiaAPL} and $C_{\gamma1(2)}=e^{2}\gamma_{1(2)}$, \cite{Luryi} we can describe the
charge redistribution as follows:
\begin{equation}
n_{\mathrm{G}}+\frac{C_{\mathrm{s}1}+C_{\mathrm{s}2}}{e^{2}}\varepsilon
_{\mathrm{F}}=\frac{1}{1+C_{\gamma2}/C_{\mathrm{c}2}}\frac{C_{\mathrm{ox}}}%
{e}\left(  V_{\mathrm{g}}-V_{\mathrm{D}}\right).
\end{equation}
Here, $C_{\mathrm{s}1(2)}\equiv(1/C_{\mathrm{c}1(2)}+1/C_{\gamma1(2)})^{-1}$
is the series capacitance of $C_{\mathrm{c}1(2)}$ and $C_{\gamma1(2)}$. The
corresponding equivalent circuit is shown in Fig.~2(b). \cite{SM}  While Eq. (1)
is a relation that generally holds for 2D systems, in graphene it has a non-trivial consequence
because $\varepsilon_\mathrm{F}$ is not proportional to $n_{\mathrm{G}}$. That is,
$C_{\mathrm{q}}$ in the equivalent circuit is a function of $n_{\mathrm{G}}$,
which leads to a nonlinear relation between $n_{\mathrm{G}}$ and
$V_{\mathrm{g}}$.

We calculated $n_{\mathrm{G}}$ vs.~$V_{\mathrm{g}}$ by solving Eq.~(1) with
$\varepsilon_{\mathrm{F}}=\hbar v\sqrt{\pi n_{\mathrm{G}}}$ ($v=1.15\times10^{6}%
$~m/s \cite{Ponomarenko}), $d_{1}=d_{2}=0.3$ nm, and $(\varepsilon_{1},\varepsilon_{2}%
)=(\varepsilon_{0},3\varepsilon_{0})$ ($\varepsilon_{0}$: vacuum permittivity). \cite{calcFootnote1} As shown by the dashed line in
Fig.~1(c), the data can be fitted well with physically reasonable values,
e.g., $\gamma_{1}=5.0\times10^{12}$ and $\gamma_{2}=3.7$~$(4.8)\times10^{13}%
$~eV$^{-1}$cm$^{-2}$ for $V_{\mathrm{g}}-V_{\mathrm{D}}>0$ ($<0$). \cite{footnoteSign} Here, we
used the same $\gamma_{1}$ value as in Refs.~\onlinecite{Janssen} and \onlinecite{Kopilov}. We emphasize that the
model in Ref.~\onlinecite{Kopilov}, which assumes only interface states below graphene, cannot
account for our results. A calculation with $\gamma_{2}=0$ and $\gamma
_{1}\rightarrow\infty$ yields a $V_{\mathrm{g}}$ dependence much larger than
that observed [dash-dotted line in Fig.~1(c)]. \cite{footnoteA}

Equation (1) demonstrates that the linear relation between $\varepsilon
_{\mathrm{F}}$ and $V_{\mathrm{g}}-V_{\mathrm{D}}$, which usually does not
hold for transport spectroscopy of graphene, does appear when the second term
on the left-hand side dominates the first term. For the $\gamma_{1}$ and
$\gamma_{2}$ values deduced above, we find this condition to be met when
$\left\vert n_{\mathrm{G}}\right\vert \ll10^{13}$~cm$^{-2}$. Consequently,
measurements in high $B$ provide the spectroscopy of the unique LL structure characteristic of Dirac electrons in graphene, as
demonstrated below.

\subsection{Results in high magnetic fields}

Figure 3(a) shows a color-scale plot of $R_{xx}$ measured as a function of
$V_{\mathrm{g}}$ and $B$. $R_{xx}$ peaks at $\left\vert \nu\right\vert >2$
separating adjacent QH regions appear as sets of unequally spaced parabolic
curves centered on the Dirac point and emanating from $B=0$. 
[For comparison, we show the
data for exfoliated graphene in Fig.~3(b).]
The white dashed lines in Fig.~3(a) represent the positions at which the four-fold degenerate
graphene LLs become half filled, i.e., $\nu=4i$ ($i\in%
\mathbb{Z}
$), calculated from $\gamma_{1}$ and $\gamma_{2}$ deduced from the
low-field data in Fig.~1(c). As the agreement between the experiment and
the calculation shows, our model remains valid in high fields and captures the
essence of the gate- and field-induced charge redistribution. In Fig.~3(f),
the positions of the $R_{xx}$ peaks in $V_{\mathrm{g}}$, measured from
$V_{\mathrm{D}}$, are plotted for the $N=1$ and $N=2$ LLs with the abscissa
scaled to $\sqrt{NB}$. The observed relation $V_{\mathrm{g}%
}-V_{\mathrm{D}}\varpropto\sqrt{NB}$ confirms that the mapping
of $R_{xx}$ peaks vs.~$V_{\mathrm{g}}$ and $B$ represents the structure of
relativistic graphene LLs.

As we show below, a more profound consequence of the relation $V_{\mathrm{g}%
}-V_{\mathrm{D}}\varpropto\varepsilon_{\mathrm{F}}$ is that, when the Fermi
level is located at one of the LLs, $V_{\mathrm{g}}-V_{\mathrm{D}}$
serves as a measure of the energy position of
that LL not only at its center but also at its tail.
 By substituting $\varepsilon_{\mathrm{F}}$ in Eq.~(1) with
the energy of the graphene LL, $E_{N}(B)=v\sqrt{2N\hbar eB}$, where $N$ is the
orbital index of the highest occupied level, we are able to calculate
$n_{\mathrm{G}}$ and hence $\nu=n_{\mathrm{G}}h/eB$ ($h$: Planck's constant)
as a function of $V_{\mathrm{g}}$ and $B$ without unknown parameters.
Figure~3(c) depicts the result calculated using $\gamma_{1}$ and
$\gamma_{2}$ appropriate to our sample. In the diagram, regions with
integer (non-integer) fillings are shown in gray (white) and the positions of
half fillings ($\nu=4N$) by dashed lines. The $V_{\mathrm{g}}$-$B$ plane is filled mostly with integer-$\nu$ regions, and
non-integer $\nu$ appears only in narrow regions around half fillings. This
happens because the Fermi level can lie in the gap between LLs by virtue of
the interface states, whose DOS exists in the LL gap. The impact of the interface states is made clearer by
comparing Fig.~3(c) with the calculations performed with smaller $\gamma_{2}$
[Figs.~3(d) and 3(e)]. As $\gamma_{2}$ increases, the LL fan diagram becomes
more nonlinear and, at the same time, integer regions become increasingly
dominant over non-integer regions.

Our model accounts for the main feature of the experimental results
sufficiently well. Yet, it is noticeable that at low $B$ the measured $R_{xx}$
peaks are broader than the calculated widths of the non-integer regions. This
is because the model does not take into account the LL\ broadening caused by
disorder. This, in turn, allows us to extract the energy width $\Gamma$ of the
extended states from the widths of the $R_{xx}$ peaks. 
Note that $\Gamma$ is otherwise not accessible through other conventional techniques including capacitance, \cite{Ponomarenko} infrared, \cite{infrared} and scanning tunneling \cite{Andrei} spectroscopy, which do not distinguish extended and localized states. \cite{footnoteTutuc}
When the overlap with adjacent LLs is negligible, the full
width at half maximum (FWHM) $\Delta V_{\mathrm{g}}$ of an $R_{xx}$ peak is
shown to be related to $\Gamma$ as
\begin{equation}
\Gamma=\frac{C_{\mathrm{ox}}}{C_{\mathrm{s}1}+C_{\mathrm{s}2}}\frac
{e}{1+C_{\gamma2}/C_{\mathrm{c}2}}\left(  \Delta V_{\mathrm{g}}-f\alpha
\cdot\frac{4eB}{h}\right).
\end{equation}
Here, $f$ is the fraction of the extended state with respect to the whole LL
DOS\ and $\alpha$ is the fraction of the LL DOS within its FWHM ($\alpha
=0.761$ for Gaussian and $\alpha=0.5$ for Lorentzian). \cite{SM} In Fig.~3(g), we
plot the measured $\Delta V_{\mathrm{g}}$ and $\Gamma$ deduced for the
$N=0$ and $N=1$ LLs assuming different $f$ values as a function of
$B$. \cite{GaussOrLorentz} 
Our analysis yields $\Gamma\approx16\pm2$ meV for the $N=0$ LL with reasonable accuracy at low $B$, even without any knowledge of the exact $f$ or $\alpha$ values. Although $\Gamma$ for the $N=1$ LL is contingent on the stronger overlap with the $N = 2$ LL, we can say that $\Gamma$'s for the $N = 0$ LL and $N = 1$ LL at low $B$ are of similar magnitude. 
This allows for an interesting contrast to be made with our results and the activation gap measurements reported for exfoliated graphene, the latter suggesting that the $N=0$ LL is much sharper than higher LLs. \cite{Giesbers} We may also make a nice comparison with theory predicting
that the $N = 0$ and other LLs exhibit different behavior
for certain types of disorder. \cite{Nomura,Katsnelson, Kawarabayashi}

As shown in Fig.~4(b), the $\nu=2$ QH effect persists up to unexpectedly high
fields, a feature commonly observed in epitaxial graphene with low carrier
densities. \cite{Jobst, Janssen, Lara-Avila, Wu, ShenJAP, Jouault} The effect has recently been discussed by Janssen \textit{et
al}. in terms of $B$-induced charge transfer from the dangling-bond states to
graphene.~\cite{Janssen} As we have demonstrated in Figs.~3(a) and 3(c), when the
interface state DOS is high, $R_{xx}$ mapping vs.~$V_{\mathrm{g}}$ and $B$
mimics the graphene LL structure. This intuitively shows that the anomalously wide $\nu=2$ QH region is a consequence
of the special property of the $N=0$ LL, namely, that it has zero energy
independent of $B$. With our model, the critical field at which the Fermi
level enters the $N=0$ LL is given by%
\begin{equation}
B_{\mathrm{c}}=\frac{h}{2e}\left[  n_{\mathrm{G}}^{0}+(C_{\mathrm{s}%
1}+C_{\mathrm{s}2})\frac{\hbar v\sqrt{\pi n_{\mathrm{G}}^{0}}}{e^{2}}\right]
\end{equation}
as a function of the zero-field carrier density $n_{\mathrm{G}}^{0}$. \cite{SM}
The second term represents the effect of
$B$-induced charge transfer. Equation~(3) reveals that interface states
both above and below the graphene layer contribute to the $B$-induced charge
transfer, which increases $B_{\mathrm{c}}$ irrespective of whether they are donors
or acceptors. Equation (3) also allows us to analyze all reported data on the
anomalously wide $\nu=2$ QH state, including ours, in a single framework by
plotting them vs.~$n_{\mathrm{G}}^{0}$.

Figure~4(a) summarizes the range over which the $\nu=2$ plateau was extended
toward high $B$ in previous reports, \cite{Jobst, Janssen, Lara-Avila, Wu, ShenJAP, Jouault} overlaid on our data replotted
vs.~$n_{\mathrm{G}}^{0}$. The figure also shows Eq.~(3) calculated with
$\gamma_{2}=0$ and different $\gamma_{1}$ values, with the limit $\gamma
_{1}\rightarrow\infty$ shown by the dashed line. The calculation with $\gamma
$'s for our sample is shown by the dash-dotted lines. It must be noted that the onset
of transport in the $N=0$ LL was not observed in any of the previous
reports; the high-$B$ end shown in Fig.~4(a) only indicates the maximum field
used in the measurements. These fields, which fall within the range from
$\gamma_{1}=\gamma_{\mathrm{d}}$ to $10\gamma_{\mathrm{d}}$ ($\gamma
_{\mathrm{d}}\equiv5\times10^{12}$~eV$^{-1}$cm$^{-2}$) if $\gamma_{2}=0$ is
assumed, thus gives only a lower bound for $\gamma_{1}$. More importantly,
as Eq.~(3) shows, any type of doping that modifies the carrier density in
graphene at $B=0$ can result in $B$-induced charge transfer in high $B$ and
thus increase $B_{\mathrm{c}}$. It is therefore possible that molecules or a gate insulator used to reduce the $B=0$ density in these studies \cite{Jobst, Janssen, Lara-Avila, Wu, ShenJAP, Jouault, Coletti, Lara-AvilaPRL,so_far} is the main cause
of the anomalously wide $\nu=2$ region. Since the contributions of $\gamma
_{1}$ and $\gamma_{2}$ are not distinguishable from $B$-sweep measurements as
evident from Eq.~(3), it is essential to examine $V_{\mathrm{g}}$ dependence,
as we have demonstrated here, to identify the exact origin of the anomalously
wide $\nu=2$ QH region and to predict the onset of transport in the $N=0$ LL.

\section{CONCLUSION}

In summary, we have demonstrated that the graphene quantum capacitance can
have an impact on transport spectroscopy through
the interplay with the interface state DOS. With high interface
state densities, gate-sweep transport measurements serve as energy
spectroscopy of graphene. Our model and analysis will be useful for
understanding various transport spectroscopy results for gated graphene devices, not
limited to epitaxial graphene but also including exfoliated graphene and their nano- and heterostructures. \cite{Tutuc}

We thank M.~Ueki for help with device fabrication. This work received support from Grants-in-Aid for Scientific Research (Nos.~21246006) from the
Ministry of Education, Culture, Sports, Science and Technology of Japan.

\appendix*
\section{Derivation of the model}

\subsection{Charge equilibrium in a gated graphene device containing interface
states}

We first derive an equation describing the charge equilibrium in a gated
graphene device in the presence of interface states both above and below the
graphene that act as charge reservoirs. For epitaxial graphene, the interface
states above and below the graphene correspond to the interface states in the
gate insulator and the dangling-bond states at the SiC, respectively.
We focus on a case where the charge
carriers are electrons. The model can be easily generalized to a case where
the charge carriers are holes.

Our model assumes two kinds of interface states with a constant density of
states $\gamma_{1}$ and $\gamma_{2}$ located below and above graphene at
distances of $d_{1}$ and $d_{2}$, respectively. [Hereafter we use subscript
$1$ ($2$) to specify quantities associated with the lower (upper) interface
states.] These interface states are capacitively coupled with graphene via the
geometrical capacitance $C_{\mathrm{c}1(2)}=\varepsilon_{1(2)}/d_{1(2)}$,
where $\varepsilon_{1(2)}$ is the effective permittivity between the lower
(upper) interface and the graphene, and thus act as charge reservoirs. Due to
the work-function differences $A_{1}$ and $A_{2}$ between graphene and the
interface states, charges are redistributed among them until their
electrochemical potentials equilibrate, where the interface states act as
donors or acceptors depending on the signs of $A_{1}$ and $A_{2}$. Gating,
which can be either via the field effect or molecular/polymer doping, induces
the charges to redistribute while maintaining the equilibrium among the
electrochemical potentials of the interface states and graphene.

We express the equilibrium charge densities in the lower interface states,
graphene, upper interface states, and gate as $-en_{1}$, $-en_{\mathrm{G}}$,
$-en_{2}$, and $-en_{\mathrm{gate}}$, respectively, where $e$ ($>0$) is the
elementary charge. Charge conservation requires that%
\begin{equation}
n_{1}+n_{\mathrm{G}}+n_{2}+n_{\mathrm{gate}}=0\tag{A1}%
\end{equation}
The equilibrium between the electrochemical potentials of the lower interface
states and graphene is expressed as%
\begin{equation}
\left(  A_{1}+\frac{n_{1}}{\gamma_{1}}\right)  +\frac{d_{1}e^{2}}%
{\varepsilon_{1}}n_{1}=\varepsilon_{\mathrm{F}},\tag{A2}%
\end{equation}
where $\varepsilon_{\mathrm{F}}$ is the Fermi level of graphene. The second
term on the left-hand side represents the change in the chemical potential of
the interface states caused by the addition or removal of electrons. (That is,
the sum of the two terms in the parenthesis represents the chemical potential
of the lower interface states.) The third term represents the electrostatic
potential difference between the lower interface states and graphene resulting
from the transfer of charge $-en_{1}$ through the plane between them (i.e.,
Gauss's law.) A similar relation holds for the upper interface states and
graphene%
\begin{equation}
\left(  A_{2}+\frac{n_{2}}{\gamma_{2}}\right)  +\frac{d_{2}e^{2}}%
{\varepsilon_{2}}\left(  n_{2}+n_{\mathrm{gate}}\right)  =\varepsilon
_{\mathrm{F}},\tag{A3}%
\end{equation}
where the terms describing the electrostatic potential difference between the
upper interface states and graphene result from the transfer of the charge
$-e(n_{2}+n_{\mathrm{gate}})$ through the plane between them.

By introducing the quantum capacitances $C_{\gamma1(2)}$ $\equiv$ $e^{2}%
\gamma_{1(2)}$ of the interface states along with the geometrical capacitances
$C_{\mathrm{c}1(2)}=\varepsilon_{1(2)}/d_{1(2)}$, we are able to deal with the
effects of the chemical potentials and the electrostatic potentials on an
equal footing and rewrite Eqs.~(A2) and (A3) in the following more intuitive
forms%
\begin{subequations}
\begin{gather}
A_{1}+\left(  \frac{1}{C_{\mathrm{c}1}}+\frac{1}{C_{\gamma1}}\right)
e^{2}n_{1}=\varepsilon_{\mathrm{F}},\tag{A4}\\
A_{2}+\left(  \frac{1}{C_{\mathrm{c}2}}+\frac{1}{C_{\gamma2}}\right)
e^{2}n_{2}+\frac{e^{2}}{C_{\mathrm{c}2}}n_{\mathrm{gate}}=\varepsilon
_{\mathrm{F}}.\tag{A5}%
\end{gather}
By eliminating $n_{1}$ and $n_{2}$ from Eqs. (A1), (A4), and (A5), we obtain
the relation between $n_{\mathrm{gate}}$ and $n_{\mathrm{G}}$%
\end{subequations}
\begin{equation}
n_{\mathrm{G}}+\frac{C_{\mathrm{s}1}+C_{\mathrm{s}2}}{e^{2}}\varepsilon
_{\mathrm{F}}=-\frac{n_{\mathrm{gate}}}{1+C_{\gamma2}/C_{\mathrm{c}2}}%
+\frac{C_{\mathrm{s}1}A_{1}+C_{\mathrm{s}2}A_{2}}{e^{2}},\tag{A6}%
\end{equation}
where $C_{\mathrm{s}1(2)}\equiv(1/C_{\mathrm{c}1(2)}+1/C_{\gamma1(2)})^{-1}$
is the series capacitance of $C_{\mathrm{c}1(2)}$ and $C_{\gamma1(2)}$. By
substituting $\varepsilon_{\mathrm{F}}=\hbar v\sqrt{\pi n_{\mathrm{G}}}$
($\hbar$: Planck's constant divided by $2\pi$, $v$: the Fermi velocity) into
Eq.~(A6), we obtain $n_{\mathrm{G}}$ as a function of $n_{\mathrm{gate}}$.

\subsection{Gate-voltage dependence of carrier density}

With field-effect gating, $n_{\mathrm{gate}}$ is related to the voltage
$V_{\mathrm{g}}$ applied to the gate. As explained later, the approximate
relation $n_{\mathrm{gate}}\approx-C_{\mathrm{ox}}V_{\mathrm{g}}/e$ holds,
where $C_{\mathrm{ox}}$ is the capacitance of the gate insulator. Thus, by
substituting $n_{\mathrm{gate}}\approx-C_{\mathrm{ox}}V_{\mathrm{g}}/e$
together with $\varepsilon_{\mathrm{F}}=\hbar v\sqrt{\pi n_{\mathrm{G}}}$ into
Eq.~(A6), we obtain $n_{\mathrm{G}}$ as a function of $V_{\mathrm{g}}$.

By setting $n_{\mathrm{G}}=0$ and $\varepsilon_{\mathrm{F}}=0$ in Eq.~(A6), we
find that the gate voltage $V_{\mathrm{D}}$ at the Dirac point can be given in
the form%
\begin{equation}
V_{\mathrm{D}}\approx-\left(  1+\frac{C_{\gamma2}}{C_{\mathrm{c}2}}\right)
\frac{C_{\mathrm{s}1}A_{1}+C_{\mathrm{s}2}A_{2}}{eC_{\mathrm{ox}}}.\tag{A7}%
\end{equation}
Hence, the relation between $V_{\mathrm{g}}$ and $n_{\mathrm{G}}$ is expressed
as follows,%
\begin{equation}
\frac{C_{\mathrm{ox}}}{e}\left(  V_{\mathrm{g}}-V_{\mathrm{D}}\right)
\approx\frac{C_{\gamma2}}{C_{\mathrm{s}2}}\left[  n_{\mathrm{G}}+\left(
C_{\mathrm{s}1}+C_{\mathrm{s}2}\right)  \frac{\hbar v\sqrt{\pi n_{\mathrm{G}}%
}}{e^{2}}\right]  .\tag{A8}%
\end{equation}
It should be noted that the parabolic dependence $\left\vert n_{\mathrm{G}%
}\right\vert \propto(V_{\mathrm{g}}-V_{\mathrm{D}})^{2}$ occurs when the
second term on the right-hand side dominates the first term. Note also that
the effects of the upper and lower interface states are not symmetric in the
equation. As explained below, this reflects the location of the interface
states with respect to graphene and the gate.

\subsection{Equivalent-circuit model}

The interplay between the interface states and graphene can be represented in
a more transparent and intuitive manner with the help of an equivalent-circuit
model. The equivalent-circuit model can be constructed by considering the
differential relations among the variations of the charge densities in the
interface states and graphene, $-e\mathrm{d}n_{1}$, $-e\mathrm{d}n_{2}$, and
$-e\mathrm{d}n_{\mathrm{G}}$, induced by a small change in the gate charge
$-e\mathrm{d}n_{\mathrm{gate}}$. By taking the derivatives of Eqs.~(A4) and
(A5) and using the relation $1/C_{\mathrm{q}}=(1/e^{2})\mathrm{d}%
\varepsilon_{\mathrm{F}}/\mathrm{d}n_{\mathrm{G}}$, we have%
\begin{align}
\frac{1}{C_{\mathrm{s}1}}e\mathrm{d}n_{1} &  =\frac{1}{C_{\mathrm{q}}%
}e\mathrm{d}n_{\mathrm{G}}\text{ (}\equiv V_{1}\text{),}\tag{A9}\\
\frac{1}{C_{\gamma2}}e\mathrm{d}n_{2} &  =\frac{1}{C_{\mathrm{q}}}%
e\mathrm{d}n_{\mathrm{G}}+\frac{1}{C_{\mathrm{c}2}}e\left(  \mathrm{d}%
n_{1}+\mathrm{d}n_{\mathrm{G}}\right)  \text{ (}\equiv V_{2}\text{).}\tag{A10}%
\end{align}
When deriving Eq.~(A10), we used the charge conservation [Eq.~(A1)]. Eq.~(A9)
implies that capacitors with capacitances $C_{\mathrm{s}1}$ and $C_{\mathrm{q}%
}$ connected in parallel hold charges $e\mathrm{d}n_{1}$ and $e\mathrm{d}%
n_{\mathrm{G}}$, respectively, at a common bias voltage $V_{1}$. Eq.~(A10)
indicates that the capacitor with $C_{\gamma2}$ holding charge $e\mathrm{d}%
n_{2}$ is connected in parallel to the series capacitors with $C_{\mathrm{q}}$
and $C_{\mathrm{c}2}$ holding charges $e\mathrm{d}n_{\mathrm{G}}$ and
$e\left(  \mathrm{d}n_{1}+\mathrm{d}n_{\mathrm{G}}\right)  $, respectively,
while sharing a common bias voltage $V_{2}$. With these considerations, we are
able to deduce the equivalent circuit as shown in Fig. 5.

\begin{figure}[htbp]
\includegraphics[ width=\linewidth ]{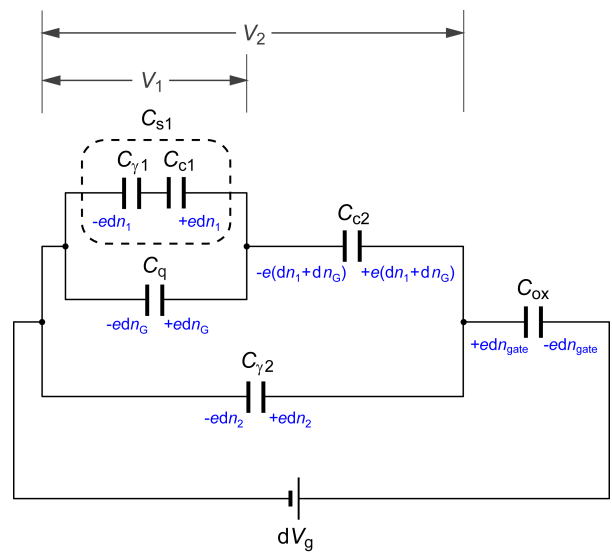} 
\caption{(Color online) Equivalent circuit for the differential relations in Eqs.~(A9) and
(A10). The model describes the changes in the amount of charge stored in each
capacitor induced by a small change in the gate voltage.}
\end{figure}

The equivalent-circuit model also tells us that the total gate capacitance
$C_{\mathrm{tot}}$, which relates the applied gate voltage and the induced
gate charge via $\mathrm{d}n_{\mathrm{gate}}/\mathrm{d}V_{\mathrm{g}%
}=-C_{\mathrm{tot}}/e$, is given by%
\begin{equation}
\frac{1}{C_{\mathrm{tot}}}=\frac{1}{C_{\mathrm{ox}}}+\frac{1}{C_{\gamma
2}+\left(  \frac{1}{C_{\mathrm{c}2}}+\frac{1}{C_{\mathrm{q}}+C_{\mathrm{s}1}%
}\right)  ^{-1}}\tag{A11}%
\end{equation}
In the absence of interface states, the above equation reduces to
$C_{\mathrm{tot}}{}^{-1}=C_{\mathrm{ox}}^{-1}+C_{\mathrm{q}}^{-1}$, which is
the relation that applies to usual cases. On the other hand, if the interface
state densities are very high such that $C_{\mathrm{ox}}\ll C_{\gamma1}$,
$C_{\gamma2}$ as demonstrated in our experiments, Eq.~(A11) reduces to
$C_{\mathrm{tot}}\approx C_{\mathrm{ox}}$.

\subsection{Landau-level filling in high magnetic fields}

In this section, we derive equations that represent the way in which the
Landau-level (LL) filling in graphene varies as a function of $V_{\mathrm{g}}$
and magnetic field $B$ in the presence of interface states. As shown in the
previous section, in zero field, $n_{\mathrm{G}}$ and $V_{\mathrm{g}}$ follow
the relation,%
\begin{equation}
\frac{C_{\mathrm{ox}}}{e}\left(  V_{\mathrm{g}}-V_{\mathrm{D}}\right)
=\frac{C_{\gamma2}}{C_{\mathrm{s}2}}\left[  n_{\mathrm{G}}+\left(
C_{\mathrm{s}1}+C_{\mathrm{s}2}\right)  \frac{\varepsilon_{\mathrm{F}}}{e^{2}%
}\right]  .\tag{A12}%
\end{equation}
In high fields, the energy spectrum of graphene is quantized into discrete LLs
whose energies are given by $E_{N}(B)=v\sqrt{2N\hbar eB}$, where $N$ ($=0$,
$1$, $2$\ldots) is the LL orbital index. When the Fermi level is located at
one of the LLs, the Fermi energy is just the energy of the highest occupied
LL, and thus $\varepsilon_{\mathrm{F}}$ in Eq.~(A12) is given as a function of
$B$ as $\varepsilon_{\mathrm{F}}=E_{N}(B)=v\sqrt{2N\hbar eB}$, with $N$ being
the orbital index of the highest occupied LL. Noting the relation
$n_{\mathrm{G}}=\nu eB/h$ ($h$: Planck's constant) and inserting
$\varepsilon_{\mathrm{F}}$ and $n_{\mathrm{G}}$ into Eq.~(A12), we obtain the
relation for $V_{\mathrm{g}}$, $\nu$, $B$, and $N$ as follows,%
\begin{equation}
\frac{C_{\mathrm{ox}}}{e}\left(  V_{\mathrm{g}}-V_{\mathrm{D}}\right)
=\frac{C_{\gamma2}}{C_{\mathrm{s}2}}\left[  \frac{\nu eB}{h}+\left(
C_{\mathrm{s}1}+C_{\mathrm{s}2}\right)  \frac{v\sqrt{2N\hbar eB}}{e^{2}%
}\right]  .\tag{A13}%
\end{equation}

When the Fermi level resides in the $N$th LL, the $\nu$ value spans a range
from $4(N-1/2)$ to $4(N+1/2)$ as $B$ or $V_{\mathrm{g}}$ is varied, where the
factor $4$ arises from the spin and valley degeneracy of graphene LLs and the
term $1/2$ from Berry's phase. Thus, by substituting $\nu=4(N-1/2)$ or $4(N+1/2)$ into Eq.~(A13) and solving it to obtain $B$ for each
value of $V_{\mathrm{g}}$, we are able to map the region where the $N$th LL is
partially filled. A result of such calculations is shown in Fig. 6. The
white regions are those where the highest occupied LL is partially filled,
i.e., where $\nu$ is non-integer. As the figure demonstrates, these regions
with non-integer $\nu$ are separated by broad regions corresponding to integer
$\nu$ (shown in gray). As discussed in the main text, these regions emerge as
a consequence of the large interface state densities.

\begin{figure}[htbp]
\includegraphics[ width=\linewidth ]{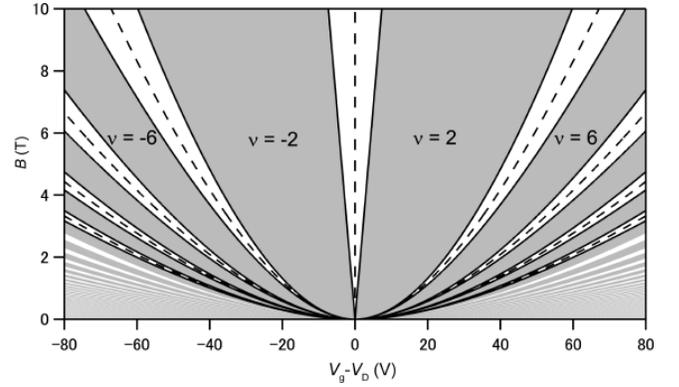} 
\caption{Mapping of regions with integer (gray) and non-integer (white)
Landau level fillings as a function of $V_{\mathrm{g}}$ and $B$, calculated
with $\gamma_{1}=5.0\times10^{12}$~eV$^{-1}$cm$^{-2}$ and $\gamma
_{2}=3.7\times10^{13}$ eV$^{-1}$cm$^{-2}$. The dashed lines represent half
fillings of four-fold degenerate graphene Landau levels.}
\end{figure}

The dashed line along the center of each white region represents the point
where the $N$th LL becomes exactly half filled, and this is obtained by
solving Eq.~(A13) with $\nu=4N$. Note that these dashed lines, which would
correspond to peaks in $R_{xx}$, are unequally spaced in $V_{\mathrm{g}}$,
with the distance increasing with decreasing $|N|$. Consequently, at each $B$,
the $\nu=2$ regions are much wider in $V_{\mathrm{g}}$ than any other integer
region. This results from the fact that graphene LLs are unequally spaced.
When the interface state densities are very high, the second term in Eq.~(A13)
dominates the first term, leading to $\left\vert V_{\mathrm{g}}-V_{\mathrm{D}%
}\right\vert \propto\sqrt{NB}$ for $|N|\geq1$. As a result, the LL fan diagram
observed in the mapping of $R_{xx}$ on the $V_{\mathrm{g}}$-$B$ plane mimics
the energy diagram of graphene LLs plotted versus $B$.

A noteworthy exception to the above argument occurs for the $N=0$ LL. Because
of its zero energy, the second term in Eq.~(A13) vanishes, leaving a linear
relation between $V_{\mathrm{g}}$ and $B$. As a result, the line defining the
high-field end of the $\nu=2$ region appears as a straight line on the
$V_{\mathrm{g}}$-$B$ plane, in marked contrast to other lines separating
integer and non-integer regions that are all curved. As detailed below, this
unique property of the $N=0$ LL is responsible for the anomalously wide
$\nu=2$ region often seen in the magnetic-field-sweep data of epitaxial graphene.

\subsection{Comparison with experiments with fixed carrier densities and onset
of transport in the $N=0$ LL}

The above discussion has focused on how the LL filling in graphene varies as a
function of $V_{\mathrm{g}}$ and $B$ in the presence of interface states.
However, in many cases, experiments on epitaxial graphene in high magnetic
fields employ samples with static gating, such as polymer or molecular gating.
The above formulation derived as a function of $V_{\mathrm{g}}$ is not
directly applicable to these experiments. To allow experiments with different
gating schemes to be compared in the same framework, in the following, we
derive a formula that does not contain $V_{\mathrm{g}}$ and, instead, contains
the carrier density in zero field ($n_{\mathrm{G}}^{0}$) as the input parameter.

As mentioned above, $C_{\mathrm{tot}}$ can be regarded as a constant that does
not depend on $B$. Accordingly, for a given value of $V_{\mathrm{g}}$, the
left-hand side of Eqs.~(A12) and (A13) can be regarded as a constant
independent of $B$. This allows us to eliminate $V_{\mathrm{g}}$ by
subtracting Eq.~(A12) for $B=0$ from Eq.~(A13) for a finite $B$. We obtain the
following equation,%
\begin{equation}
\frac{e^{2}}{C_{\mathrm{s}1}+C_{\mathrm{s}2}}\left[  \frac{\nu eB}%
{h}-n_{\mathrm{G}}^{0}\right]  +\sqrt{\pi}\hbar v\left[  \sqrt{\frac{4NeB}{h}%
}-\sqrt{n_{\mathrm{G}}^{0}}\right]  =0,\tag{A14}%
\end{equation}
where we used the relation $\varepsilon_{\mathrm{F}}=\hbar v\sqrt{\pi
n_{\mathrm{G}}^{0}}$ that holds at $B=0$. When the interface state density is very
high, neglecting the first two terms in Eq.~(A14) yields $B=hn_{\mathrm{G}%
}^{0}/4Ne$. This equation indicates that, even when the interface state
density is very high, the magnetic field positions of $R_{xx}$ peaks at
$\nu=4N$ ($N\geq1$) follow exactly what is expected from the zero-field
carrier density.

The above argument is not applicable to the $N=0$ LL. The center of the
$R_{xx}$ peak associated with the $N=0$ LL is fixed at the charge neutrality
point and therefore is not a function of $B$. We thus take the critical field
$B_{\mathrm{c}}$ at which the Fermi level enters the $N=0$ LL as a measure
characterizing the width of the $\nu=2$ region along the $B$ axis. By setting
$\nu=2$ and $N=0$ in Eq.~(A14), $B_{\mathrm{c}}$ is obtained as a function of
$n_{\mathrm{G}}^{0}$,%
\begin{align}
B_{\mathrm{c}} &  =\frac{h}{2e}\left[  n_{\mathrm{G}}^{0}+\left(
C_{\mathrm{s}1}+C_{\mathrm{s}2}\right)  \frac{\hbar v\sqrt{\pi
n_{\mathrm{G}}^{0}}}{e^{2}}\right]  \tag{A15}\\
&  =\frac{h}{2e}\left[  n_{\mathrm{G}}^{0}+\left(  C_{\mathrm{s}%
1}+C_{\mathrm{s}2}\right)  \frac{\varepsilon_{\mathrm{F}}^{0}}{e^{2}}\right]
\nonumber
\end{align}
where $\varepsilon_{\mathrm{F}}^{0}\equiv\hbar v\sqrt{\pi n_{\mathrm{G}}^{0}}$
is the Fermi energy of graphene at $B=0$. Note that the first term of
Eq.~(A15), $hn_{\mathrm{G}}^{0}/2e$ ($\equiv B_{\mathrm{c}}^{0}$), is simply
the magnetic field at which $\nu=2$ is expected from the zero-field density.
The second term represents the effects of interface states that transfer
charged carriers to graphene as $B$ is increased, thereby pinning the filling
factor at $\nu=2$ over an unexpectedly broad range of $B$. In Eq.~(A15), the
second term can be viewed as the density of carriers stored in reservoirs with
the capacitances $C_{\mathrm{s}1}$ and $C_{\mathrm{s}2}$, which are both
biased at a voltage $\varepsilon_{\mathrm{F}}^{0}/e$ at $B=0$. The reservoirs
here are the interface states, whose capacitances are determined as series
capacitances of their quantum capacitances and the geometrical capacitances
with respect to graphene. In high magnetic fields, all of these charges are
transferred to graphene when the Fermi level enters the zero-energy $N=0$ LL,
where the \textquotedblleft bias voltage\textquotedblright\ vanishes.

It is also worth noting that, when expressed as a function of $n_{\mathrm{G}%
}^{0}$ as in Eq.~(A15), the effects of the upper and lower interface states on
$B_{\mathrm{c}}$ appear to be equivalent, and are only weighted by
$C_{\mathrm{s}1}$ and $C_{\mathrm{s}2}$. This clearly shows that the
individual contributions of the upper and lower interface states cannot be
distinguished from a $B$-dependent study for a fixed value of $n_{\mathrm{G}%
}^{0}$. This contrasts with the case when the equations are expressed as a
function of $V_{\mathrm{g}}$, where the effects of the upper and lower
interface states appear asymmetric, reflecting their asymmetric locations with
respect to the gate and graphene. This allows us for the first time to
estimate the individual contributions of the upper and lower interface states,
which highlights the significance of our $V_{\mathrm{g}}$-dependent study.

\subsection{Energy width of extended state}

Our model has so far not taken into account the broadening of LLs due to
disorder. When the overlap with adjacent LLs is negligible, effects of finite
LL widths can be incorporated in the model as follows. Suppose that the gate
voltage is swept from $V_{\mathrm{g}}$ to $V_{\mathrm{g}}+\delta
V_{\mathrm{g}}$ so that the Fermi level moves from $\varepsilon_{\mathrm{F}}$
to $\varepsilon_{\mathrm{F}}+\delta E$ within a disorder-broadened LL across
its center. The resultant change in the carrier density from $n_{\mathrm{G}}$
to $n_{\mathrm{G}}+\delta n_{\mathrm{G}}$ is related to $\delta V_{\mathrm{g}%
}$ and $\delta E$ through Eq.~(A12) as follows
\begin{equation}
\delta n_{\mathrm{G}}+\frac{C_{\mathrm{s}1}+C_{\mathrm{s}2}}{e^{2}}\delta
E=\frac{1}{1+C_{\gamma2}/C_{\mathrm{c}2}}\frac{C_{\mathrm{ox}}}{e}\delta
V_{\mathrm{g}}.\tag{A16}%
\end{equation}
If $\delta E$ corresponds to the full width at half maximum (FWHM) of the
disorder-broadened LL, $\delta n_{\mathrm{G}}$ can be expressed as $\delta
n_{\mathrm{G}}=\alpha\cdot4eB/h$, where $\alpha$ is the fraction of the LL
density of states within its FWHM [$\alpha=\operatorname{erf}(\ln
2)\approx0.761$ for Gaussian and $\alpha=1/2$ for Lorentzian] and the factor
$4$ originates from the spin and valley degeneracy. We thus obtain%
\begin{equation}
\alpha\cdot\frac{4eB}{h}+\frac{C_{\mathrm{s}1}+C_{\mathrm{s}2}}{e^{2}}\delta
E=\frac{1}{1+C_{\gamma2}/C_{\mathrm{c}2}}\frac{C_{\mathrm{ox}}}{e}\delta
V_{\mathrm{g}}.\tag{A17}%
\end{equation}
Since transport measurements probe only extended states, we introduce a parameter
$f$ that denotes the fraction of the extended states in the total LL density
of states. The width $\Delta V_{\mathrm{g}}$of the longitudinal resistance
peak can thus be related to the energy width $\Gamma$ of the extended states
as
\begin{equation}
\Gamma=\frac{C_{\mathrm{ox}}}{C_{\mathrm{s}1}+C_{\mathrm{s}2}}\frac
{e}{1+C_{\gamma2}/C_{\mathrm{c}2}}\left(  \Delta V_{\mathrm{g}}-f\alpha
\cdot\frac{4eB}{h}\right)  \tag{A18}%
\end{equation}
by replacing $\alpha$, $\delta E$, and $\delta V_{\mathrm{g}}$ in Eq.~(A17)
with $f\alpha$, $\Gamma$, and $\Delta V_{\mathrm{g}}$, respectively.

\bigskip



\begin{thebibliography}{99}





\bibitem{Novoselov} K. S. Novoselov, A. K. Geim, S. V. Morozov, D. Jiang, Y. Zhang, S. V. Dubonos, I. V. Grigorieva, and A. A. Firsov, Science \textbf{306}, 666 (2004).
\bibitem{Novoselov2} K. S. Novoselov, A. K. Geim, S. V. Morozov, D. Jiang, M. I. Katsnelson, I. V. Grigorieva, S. V. Dubonos, and A. A. Firsov, Nature \textbf{438}, 197 (2005).
\bibitem{Zhang} Y. Zhang, Y. Tan, H. L. Stormer, and P. Kim, Nature \textbf{438}, 201 (2005).

\bibitem{Chen} Z. Chen and J. Appenzeller, in IEEE International Electron Devices Meeting 2008, Technical Digest (IEEE. New York, 2008), p. 509-512. 
\bibitem{XiaNatureNanotech} J. Xia, F. Chen, J. Li, and N. Tao, Nature Nanotech. \textbf{4}, 505 (2009).
\bibitem{Ponomarenko} L. A. Ponomarenko, R. Yang, R.V. Gorbachev, P. Blake, A. S. Mayorov, K. S. Novoselov, M. I. Katsnelson, and A. K. Geim, Phys .Rev Lett, \textbf{105}, 136801 (2010).
\bibitem{Ensslin} S. Droscher, P. Roulleau, F. Molitor, P. Studerus, C. Stampfer, K. Ensslin, and T. Ihn, Appl. Phys. Lett. \textbf{96}, 152104 (2010).
\bibitem{XiaAPL} J. L. Xia, F. Chen, J. L. Tedesco, D. K. Gaskill, R. L. Myers-Ward, C. R. Eddy, D. K. Ferry, and N. J. Tao,  Appl. Phys. Lett. \textbf{96}, 162101 (2010).
\bibitem{Luryi} S. Luryi, Appl. Phys. Lett. \textbf{52}, 501 (1988).

\bibitem{Jobst} J. Jobst, D. Waldmann, F. Speck, R. Hirner, D. K. Maude, T. Seyller, and H. B. Weber, Phys. Rev. B \textbf{81}, 195434 (2010).
\bibitem{Janssen} T. J. B. M. Janssen, A. Tzalenchuk, R. Yakimova, S. Kubatkin, S. Lara-Avila, S. Kopylov, and V. I. Fal'ko, Phys. Rev. B \textbf{83}, 233402 (2011).
\bibitem{Lara-Avila} S. Lara-Avila, K. Moth-Poulsen, R. Yakimova, T. Bjornholm, V. Falfko, A. Tzalenchuk, and S. Kubatkin, Adv. Mater. \textbf{28}, 878 (2011). 

\bibitem{Wu} X Wu, Y. Hu, M Ruan, N. K. Madiomanana, J. Hankinson, M. Sprinkle, C. Berger, and W. A. de Heer, Appl. Phys. Lett. \textbf{95}, 223108 (2009).
\bibitem{ShenJAP} T. Shen, A. T. Neal, M. L. Bolen, J. J. Gu, L. W. Engel, M. A. Capano, and P. D. Ye, J. Appl. Phys. \textbf{111}, 013716 (2012).
\bibitem{Jouault} B. Jouault, N. Camara, B. Jabakhanji, A. Caboni, C. Consejo, P. Godignon, D. K. Maude, and J. Camassel, Appl. Phys. Lett. \textbf{100}, 052102 (2012).

\bibitem{Tutuc}
S. Kim, I. Jo, D. C. Dillen, D. A. Ferrer, B. Fallahazad, Z. Yao, S. K. Banerjee, and E. Tutuc, Phys. Rev. Lett. \textbf{108}, 116404 (2012).

\bibitem{TanabeAPEX} S. Tanabe, Y. Sekine, H. Kageshima, M. Nagase, and H. Hibino, Appl. Phys. Exp. \textbf{3}, 075102 (2010).
\bibitem{Kopilov} S. Kopylov, A. Tzalenchuk, S Kubatkin, and V I. Falfko, Appl. Phys. Lett. \textbf{97}, 112109 (2010).
\bibitem{Varchon}  F. Varchon, R. Feng, J. Hass, X. Li, B. N. Nguyen, C. Naud, P. Mallet, J.-Y. Veuillen, C. Berger, E. H. Conrad, L. Magaud, Phys. Rev. Lett. \textbf{99}, 126805 (2007).

\bibitem{Sonde} S. Sonde, F. Giannazzo, V. Raineri, R. Yakimova, J.-R. Huntzinger, A. Tiberj, and J. Camassel, Phys. Rev. B \textbf{80}, 241406 (R) (2009).

\bibitem{Sze}S. M. Sze, Semiconductor Devices: Physics and Technology (Wiley, New York, 2001).

\bibitem{SM} See Appendix for the derivation of the equation.

\bibitem{calcFootnote1} $d_1 = 0.3$ nm and $\varepsilon_1 = \varepsilon_0$ are assumed as in Refs. \onlinecite{Janssen} and \onlinecite{Kopilov}. 

\bibitem{footnoteSign}
The slight asymmetry between electron and hole bands is related to a hysteresis in the gate sweep. All the data presented in this paper were taken with $V_\mathrm{g}$ ramped from positive to negative. For gate sweeps in the opposite direction, the asymmetry is reversed. The hysteresis suggests that the carrier accumulation in graphene tends to be retarded when $V_\mathrm{g}$ is tuned away from the Dirac point. Although the detailed mechanism of the hysteresis is unknown, we emphasize that such dynamic behavior is beyond the scope of this paper and does not affect our conclusions on the static properties.

\bibitem{footnoteA} This can be understood by
noting that $C_{\mathrm{s}1}\rightarrow C_{\mathrm{c}1}$ for $\gamma
_{1}\rightarrow\infty$; that is, the effect of $\gamma_{1}$ is limited by
$C_{\mathrm{c}1}$, which enters in series with $C_{\gamma1}$ in the equivalent
circuit [Fig.~2(b)].

\bibitem{infrared} Z. Jiang, E. A. Henriksen, L. C. Tung, Y.-J. Wang, M. E. Schwartz, M.Y. Han, P. Kim, and H. L. Stormer, Phys. Rev. Lett. \textbf{98}, 197403 (2007).
\bibitem{Andrei} G. Li, A. Luican, and E. Y. Andrei, Phys. Rev. Lett. \textbf{102}, 176804 (2009).

\bibitem{footnoteTutuc} A similar method of probing the extended state of the
graphene LL was recently reported for a heterostructure
device using double layers of graphene. \cite{Tutuc}


\bibitem{GaussOrLorentz} Here we assumed the Gaussian function. For the Lorentzian, the
$f$ values in Fig.~3(g) are multiplied by a factor 1.52.

\bibitem{Giesbers} A. J. M. Giesbers, U. Zeitler, M. I. Katsnelson, L. A. Ponomarenko, T. M. Mohiuddin, and J. C. Maan, Phys. Rev. Lett. \textbf{99}, 206803 (2007).


\bibitem{Nomura} K. Nomura, S. Ryu, M. Koshino, C. Mudry, and A. Furusaki, Phys. Rev. Lett. \textbf{100}, 246806 (2008).
\bibitem{Katsnelson} M. I. Katsnelson, Mater. Today \textbf{10}, 20 (2007).
\bibitem{Kawarabayashi} T. Kawarabayashi, Y. Hatsugai, and H. Aoki, Phys. Rev. Lett. \textbf{103}, 156804 (2009).


\bibitem{Coletti}  C. Coletti, C. Riedl, D. S. Lee, B. Krauss, L. Patthey, K. von Klitzing, J. H. Smet, and U. Starke, Phys. Rev. B \textbf{81}, 235401 (2010).

\bibitem{Lara-AvilaPRL} S. Lara-Avila, A. Tzalenchuk, S. Kubatkin, R. Yakimova, T. J. B. M. Janssen, K. Cedergren,T. Bergsten, and V. Fal'ko, Phys. Rev. Lett. \textbf{107}, 166602 (2011) 

\bibitem{so_far} Within our model, $V_\mathrm{D}$ contains the information on the work-function difference $A_{1(2)}$ between graphene and the lower (upper) interface states (see Section 2 of Appendix.). We can thus deduce $A_1 =$ $\sim 1.0-1.3$ eV, $A_2 = -0.08 \sim -0.13$ eV for $V_\mathrm{D} = -33$ V, by applying our model to the situation before and after the gate insulator is deposited. Note that the electron density in our epitaxial graphene is reduced significantly from $\sim(3-4) \times 10^{12} \mathrm{cm}^{-2}$ to several $10^{11} \mathrm{cm}^{-2}$ simply by the deposition of an HSQ/SiO$_2$ insulator. This is similar to the methods used to reduce electron densities \cite{Jobst, Janssen, Lara-Avila} or even to improve sample quality. \cite{Lara-AvilaPRL} 








\end{thebibliography}
\end{document}